\documentclass[onecolumn,11pt]{article}
\usepackage[top=1in, bottom=1in, left=1in, right=1in]{geometry}
\usepackage{amsmath,amsfonts,amscd,amssymb}
\usepackage{graphicx}
\usepackage{hyperref}
\usepackage{url}
\usepackage{xcolor}
\usepackage{caption}
\usepackage{mathtools}
\usepackage[numbers,sort&compress]{natbib}

\usepackage[bottom,flushmargin,hang,multiple]{footmisc}
\usepackage{lipsum}
\newcommand\blfootnote[1]{%
  \begingroup
  \renewcommand\thefootnote{}\footnote{#1}%
  \addtocounter{footnote}{-1}%
  \endgroup
}

\definecolor{header1}{cmyk}{0,0,0,1}

\title{\vspace{-.65in}{\fontsize{16}{16}\selectfont \textbf{PySINDy: A comprehensive Python package\\for robust sparse system identification}}\vspace{-.15in}}

\author{
Alan A. Kaptanoglu$^{1^*}$, Brian M. de Silva$^{2*\dagger}$, Urban Fasel$^3$, Kadierdan Kaheman$^3$,\\ Andy J. Goldschmidt$^1$, Jared L. Callaham$^3$,
Charles B. Delahunt$^2$, Zachary G. Nicolaou$^2$,\\Kathleen Champion$^{2\dagger}$, Jean-Christophe Loiseau$^4$, J. Nathan Kutz$^2$, Steven L. Brunton$^3$\\
\footnotesize{$^1$ Department of Physics, University of Washington, Seattle, WA 98195, United States}\\
\footnotesize{$^2$ Department of Applied Mathematics, University of Washington, Seattle, WA 98195, United States}\\
\footnotesize{$^3$ Department of Mechanical Engineering, University of Washington, Seattle, WA 98195, United States}\\
\footnotesize{$^4$ Arts et M\'{e}tiers Institute of Technology, CNAM, DynFluid, HESAM Universit\'{e}, F-75013, Paris, France\vspace{-.2in}}
}
\date{}

\begin{document}
\maketitle
\vspace{-0.3in}

\blfootnote{$^*$ Corresponding authors (akaptano@uw.edu, briandesilva1@gmail.com).}
\blfootnote{$^\dagger$ Work performed independently of employment at Amazon.}

\section*{Summary}
Automated data-driven modeling, the process of directly discovering the governing equations of a system from data, is increasingly being used across the scientific community. \texttt{PySINDy} is a Python package that provides tools for applying the sparse identification of nonlinear dynamics (SINDy) approach to data-driven model discovery. In this major update to \texttt{PySINDy}, we implement several advanced features that enable the discovery of more general differential equations from noisy and limited data. The library of candidate terms is extended for the identification of actuated systems, partial differential equations (PDEs), and implicit differential equations. Robust formulations, including the integral form of SINDy and ensembling techniques, are also implemented to improve performance for real-world data. Finally, we provide a range of new optimization algorithms, including several sparse regression techniques and algorithms to enforce and promote inequality constraints and stability. Together, these updates enable entirely new SINDy model discovery capabilities that have not been reported in the literature, such as constrained PDE identification and ensembling with different sparse regression optimizers.

\section*{Statement of need}
Traditionally, the governing laws and equations of nature have been derived from first principles and based on rigorous experimentation and expert intuition. 
In the modern era, cheap and efficient sensors have resulted in an unprecedented growth in the availability of measurement data, opening up the opportunity to perform automated model discovery using data-driven modeling. These data-driven approaches are also increasingly useful for processing and interpreting the information in these large datasets.
A number of such approaches have been developed in recent years, including the dynamic mode decomposition~\cite{schmid2010dynamic, Kutz2016book}, Koopman theory~\cite{Brunton2021koopman}, nonlinear autoregressive algorithms~\cite{Billings2013book}, neural networks~\cite{pathak2018model,vlachas2018data,Raissi2019jcp}, Gaussian process regression~\cite{raissi2017machine}, operator inference and reduced-order modeling~\cite{Benner2015siamreview,peherstorfer2016data,qian2020lift}, genetic programming~\cite{Bongard2007pnas,schmidt_distilling_2009}, and sparse regression~\cite{brunton2016pnas}.
These approaches have seen many variants and improvements over the years, so data-driven modeling software must be regularly updated to remain useful to the scientific community. The SINDy approach has experienced particularly rapid development, motivating this major update to aggregate these innovations into a single open-source tool that is transparent and easy to use for non-experts or scientists from other fields.

The original \texttt{PySINDy} code~\cite{de2020pysindy} provided an implementation of the traditional SINDy method~\cite{brunton2016pnas}, which 
assumes that the dynamical evolution of a state variable $\mathbf{q}(t)\in\mathbb{R}^n$ follows an ODE described by a function $\mathbf{f}$,
\begin{equation}\label{eq:sindy_eq}
   \frac{d}{dt} \mathbf{q} = \mathbf{f}(\mathbf{q}).
\end{equation}
SINDy approximates the dynamical system $\mathbf{f}$ in Eq.~\eqref{eq:sindy_eq} as a sparse combination of terms from a library of candidate basis functions $\boldsymbol{\theta}(\mathbf{q}) = [\theta_1(\mathbf{q}),\theta_2(\mathbf{q}),\dots,\theta_p(\mathbf{q})]$ 
\begin{equation}\label{eq:sindy_expansion}
\mathbf{f}(\mathbf{q})\approx \sum_{k=1}^{p}\theta_k(\mathbf{q})\boldsymbol\xi_k, \quad \text{or equivalently} \quad \frac{d}{dt}\mathbf{q} \approx \mathbf{\Theta}(\mathbf{q})\mathbf{\Xi},
\end{equation}
where $\boldsymbol{\Xi} = [\boldsymbol\xi_1,\boldsymbol\xi_2,\dots,\boldsymbol\xi_p]$ contain the sparse coefficients. In order for this strategy to be successful, a reasonably accurate approximation of $\mathbf{f}(\mathbf{q})$ should exist as a sparse expansion in the span of $\boldsymbol{\theta}$. Therefore, background scientific knowledge about expected terms in $\mathbf{f}(\mathbf{q})$ can be used to choose the library $\boldsymbol{\theta}$. 
To pose SINDy as a regression problem, we assume we have a set of state measurements sampled at time steps $t_1, ..., t_m$ and rearrange the data into the data matrix $\mathbf{Q} \in \mathbb{R}^{m\times n}$, \begin{eqnarray}\label{eq:Q_matrix}
\mathbf{Q} = \begin{bmatrix}
q_1(t_1) & q_2(t_1) & \cdots & q_n(t_1)\\
q_1(t_2) & q_2(t_2) & \cdots & q_n(t_2)\\
\vdots & \vdots & \ddots & \vdots \\
q_1(t_m) & q_2(t_m) & \cdots & q_n(t_m)
\end{bmatrix}
\label{Eq:DataMatrix}.
\end{eqnarray}
A matrix of derivatives in time, $\mathbf Q_t$, is defined similarly and can be numerically computed from $\mathbf{Q}$. \texttt{PySINDy} defaults to second order finite differences for computing derivatives, although a host of more sophisticated methods are now available, including arbitrary order finite differences, Savitzky-Golay derivatives (i.e. polynomial-filtered derivatives), spectral derivatives with optional filters, arbitrary order spline derivatives, and total variational derivatives~\cite{ahnert2007numerical,chartrand2011numerical,tibshirani2011solution}.

After $\mathbf Q_t$ is obtained, Eq.~\eqref{eq:sindy_expansion} becomes $\mathbf Q_t \approx \mathbf{\Theta}(\mathbf{Q})\mathbf{\Xi}$ and the goal of the SINDy sparse regression problem is to choose a sparse set of coefficients $\mathbf{\Xi}$ that accurately fits the measured data in $\mathbf Q_t$. We can promote sparsity in the identified coefficients via a sparse regularizer $R(\mathbf{\Xi})$, such as the $l_0$ or $l_1$ norm, and use a sparse regression algorithm such as SR3~\cite{champion2020unified} to solve the resulting optimization problem,
\begin{equation}\label{eq:sindy_regression}
  \text{argmin}_{\boldsymbol\Xi}\|\mathbf Q_t - \boldsymbol\Theta(\mathbf{Q}) \boldsymbol\Xi\|^2 + R(\boldsymbol\Xi).
\end{equation}

The original \texttt{PySINDy} package was developed to identify a particular class of systems described by Eq.~\eqref{eq:sindy_eq}.
Recent variants of the SINDy method are available that address systems with control inputs and model predictive control (MPC)~\cite{Kaiser2018prsa,fasel2021sindy}, systems with physical constraints~\cite{Loiseau2017jfm,kaptanoglu2020physics}, implicit ODEs~\cite{mangan2016inferring,kaheman2020sindy}, PDEs~\cite{Rudy2017sciadv,Schaeffer2017prsa}, and weak form ODEs and PDEs~\cite{Schaeffer2017pre,Reinbold2020pre,messenger2021weakpde}. Other methods, such as ensembling and sub-sampling~\cite{maddu2019stability,reinbold2021robust,delahunt2021toolkit}, are often vital for making the identification of Eq.~\eqref{eq:sindy_eq} more robust. 
In order to incorporate these new developments and accommodate the wide variety of possible dynamical systems, we have extended \texttt{PySINDy} to a more general setting and added significant new functionality. Our code\footnote{\url{https://github.com/dynamicslab/pysindy}} is thoroughly documented, contains extensive examples, and integrates a wide range of functionality, some of which may be found in a number of other local SINDy implementations\footnote{\url{https://github.com/snagcliffs/PDE-FIND}, \url{https://github.com/eurika-kaiser/SINDY-MPC},\\ \url{https://github.com/dynamicslab/SINDy-PI}, \url{https://github.com/SchatzLabGT/SymbolicRegression},\\ \url{https://github.com/dynamicslab/databook_python}, \url{https://github.com/sheadan/SINDy-BVP},\\ \url{https://github.com/sethhirsh/BayesianSindy}, \url{https://github.com/racdale/sindyr},\\ \url{https://github.com/SciML/DataDrivenDiffEq.jl}, \url{https://github.com/MathBioCU/WSINDy_PDE},\\ \url{https://github.com/pakreinbold/PDE_Discovery_Weak_Formulation}, \url{https://github.com/ZIB-IOL/CINDy}}. In contrast to some of these existing codes, \texttt{PySINDy} is completely open-source, professionally-maintained (for instance, providing unit tests and adhering to PEP8 stylistic standards), and minimally dependent on non-standard Python packages.

\section*{New Features}
Given spatiotemporal data $\mathbf{Q}(\mathbf{x}, t) \in \mathbb{R}^{m\times n}$, and optional control inputs $\mathbf{u} \in \mathbb{R}^{m \times r}$ (note $m$ has been redefined here to be the product of the number of spatial measurements and the number of time samples), \texttt{PySINDy} can now approximate algebraic systems of PDEs (and corresponding weak forms) in an arbitrary number of spatial dimensions. Assuming the system is described by a function $\mathbf{g}$, we have
\begin{equation}\label{eq:pysindy_eq}
    \mathbf{g}(\mathbf{q},\mathbf q_t, \mathbf q_x, \mathbf q_y, \mathbf q_{xx}, ..., \mathbf{u}) = 0.
\end{equation}
ODEs, implicit ODEs, PDEs, and other dynamical systems are subsets of Eq.~\eqref{eq:pysindy_eq}. We can accommodate control terms and partial derivatives in the SINDy library by adding them as columns in $\mathbf{\Theta}(\mathbf{Q})$, which becomes $\mathbf{\Theta}(\mathbf{Q}, \mathbf Q_t, \mathbf Q_x, ..., \mathbf{u})$. 

\begin{figure}
    \centering
    \includegraphics[width=0.99\linewidth]{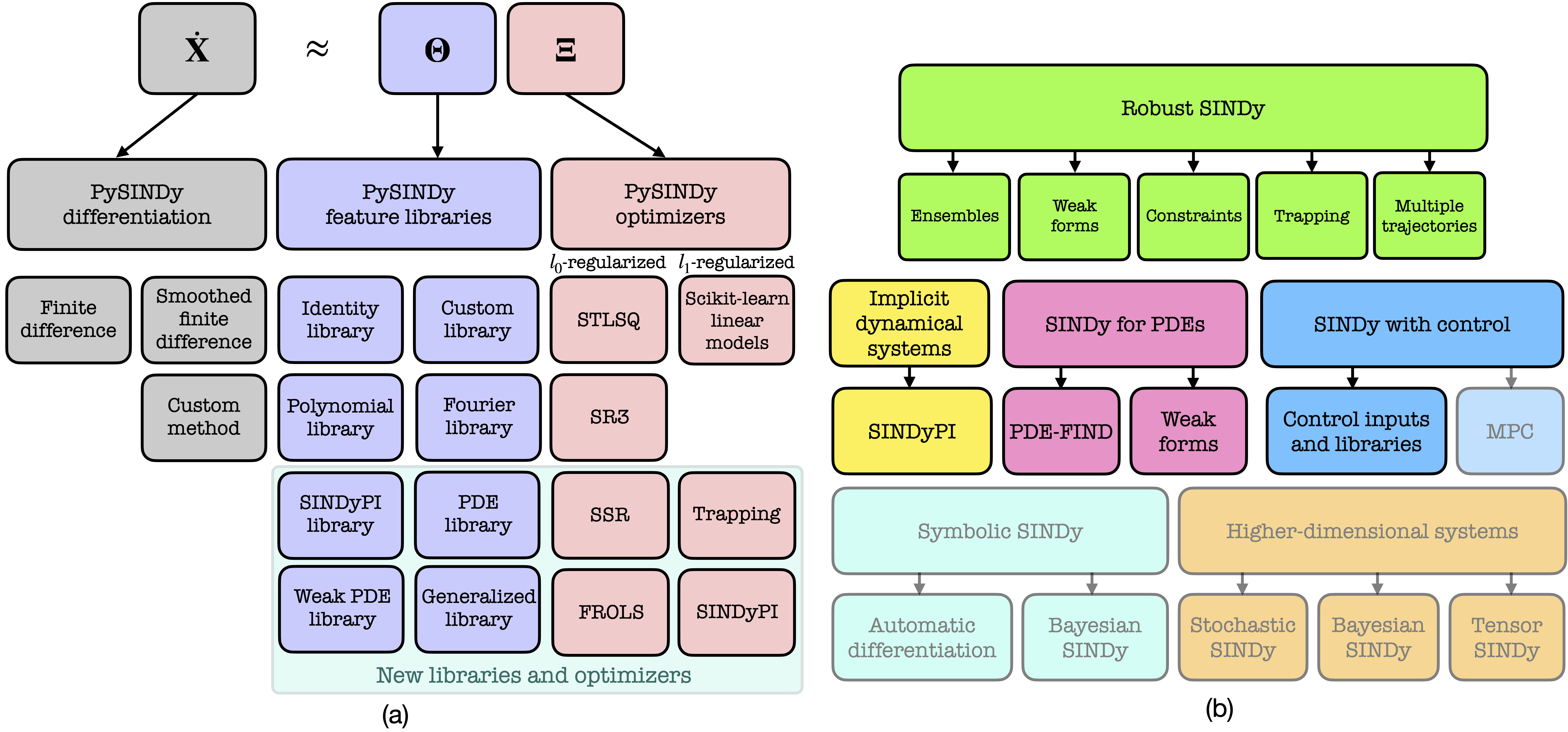}
    \caption{Summary of SINDy features organized by (a) \texttt{PySINDy} structure and (b) functionality. (a) Hierarchy from the sparse regression problem solved by SINDy, to the submodules of \texttt{PySINDy}, to the individual optimizers, libraries, and differentiation methods implemented in the code.\\ (b) Flow chart for organizing the SINDy variants and functionality in the literature. Bright color boxes indicate the features that have been implemented through this work, roughly organized by functionality. Semi-transparent boxes indicate features that have not yet been implemented.}
    \label{fig:package-structure}
    \vspace{0.15in}
\end{figure}

In addition, we have extended \texttt{PySINDy} to handle more complex modeling scenarios, including trapping SINDy for provably stable ODE models for fluids~\cite{kaptanoglu2021promoting}, models trained using multiple dynamic trajectories, and the generation of many models with sub-sampling and ensembling methods~\cite{fasel2021ensemble} for cross-validation and probabilistic system identification. In order to solve Eq.~\eqref{eq:pysindy_eq}, \texttt{PySINDy} implements several different sparse regression algorithms. Greedy sparse regression algorithms, including step-wise sparse regression (SSR)~\cite{boninsegna2018sparse} and forward regression orthogonal least squares (FROLS)~\cite{Billings2013book}, are now available. For maximally versatile candidate libraries, the new \texttt{GeneralizedLibrary} class allows for tensoring, concatenating, and otherwise combining many different candidate libraries, along with optionally specifying a subset of the inputs to use for generating each of the libraries. Fig.~\ref{fig:package-structure} illustrates the \texttt{PySINDy} code structure, changes, and high-level goals for future work, and \textcolor{blue}{\href{https://www.youtube.com/playlist?list=PLN90bHJU-JLoOfEk0KyBs2qLTV7OkMZ25}{YouTube tutorials}} for this new functionality are available online.

\begin{figure}[t]
    \centering
    \includegraphics[width=0.98\linewidth]{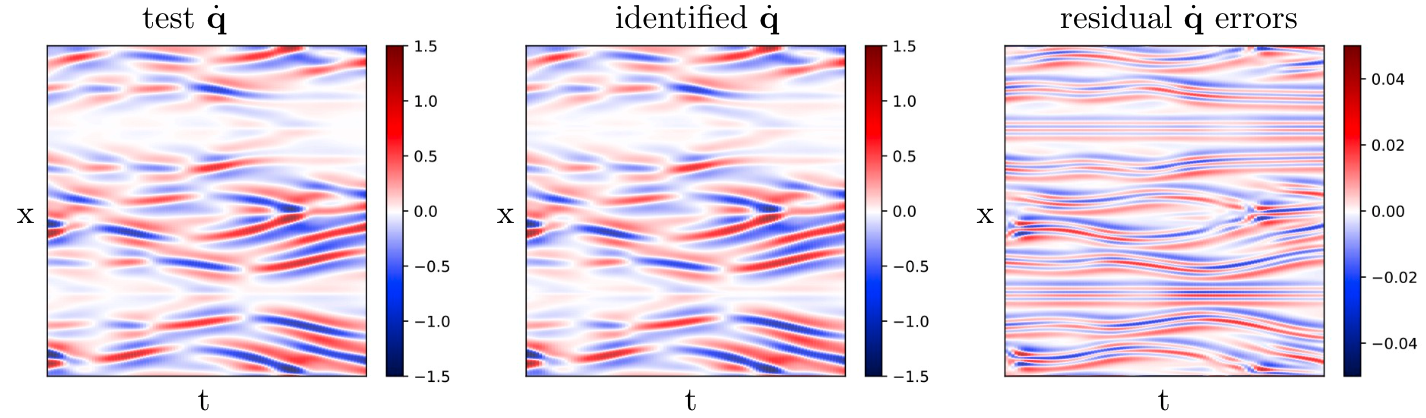}
    \caption{\texttt{PySINDy} can now be used for PDE identification; we illustrate this new capability by accurately capturing a set of testing data from the Kuramoto-Sivashinsky system, described by $q_t = -qq_x - q_{xx} - q_{xxxx}$. The identified model is $q_t = -0.98qq_x -0.99q_{xx} - 1.0q_{xxxx}$.}
    \label{fig:pde_id}
\end{figure}

\begin{figure}[t]
    \centering
    \includegraphics[width=0.98\linewidth]{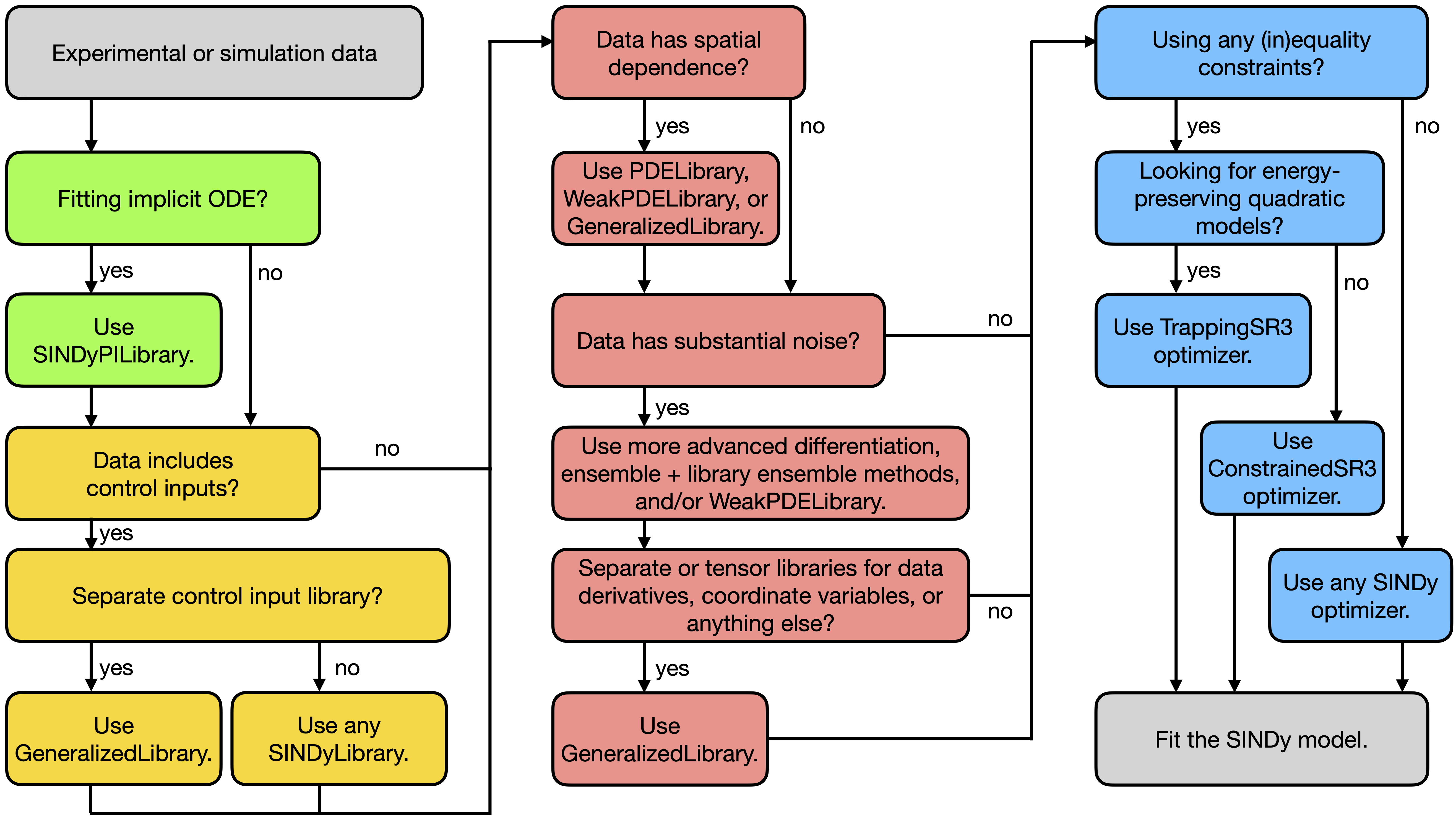}
    \caption{This flow chart summarizes how \texttt{PySINDy} users can start with a dataset and systematically choose the proper candidate library and sparse regression optimizer that are tailored for a specific scientific task. }
    \label{fig:flow_chart}
\end{figure}

\texttt{PySINDy} includes extensive Jupyter notebook tutorials that demonstrate the usage of various features of the package and reproduce nearly the entirety of the examples from the original SINDy paper~\cite{brunton2016pnas}, trapping SINDy paper~\cite{kaptanoglu2021promoting}, and the PDE-FIND paper~\cite{Rudy2017sciadv}. 
We include an extended example for the quasiperiodic shear-driven cavity flow~\cite{callaham2021role}.
As a simple illustration of the new functionality, we demonstrate how SINDy can be used to identify the Kuramoto-Sivashinsky (KS) PDE from data. We train the model on the first 60\% of the data from Rudy et al.~\cite{Rudy2017sciadv}, which in total contains 1024 spatial grid points and 251 time steps. The KS model is identified correctly and the prediction for $\dot{\mathbf{q}}$ on the remaining testing data indicates strong performance in Fig.~\ref{fig:pde_id}. Lastly, we provide a useful flow chart in Fig.~\ref{fig:flow_chart} so that users can make informed choices about which advanced methods are suitable for their datasets. 

\section*{Conclusion}
The goal of the \texttt{PySINDy} package is to enable anyone with access to measurement data to engage in scientific model discovery. The package is designed to be accessible to inexperienced users, adhere to \texttt{scikit-learn} standards, include most of the existing SINDy variations in the literature, and provide a large variety of functionality for more advanced users. We hope that researchers will use and contribute to the code in the future, pushing the boundaries of what is possible in system identification.

\section*{Acknowledgments}
\texttt{PySINDy} is a fork of \texttt{sparsereg}~\citep{markus_quade_sparsereg}. 
SLB, AAK, KK, and UF acknowledge support from the Army Research Office (ARO  W911NF-19-1-0045).
JLC acknowledges support from funding support from the Department of Defense (DoD) through the National Defense Science \& Engineering Graduate (NDSEG) Fellowship Program. ZGN is a Washington Research Foundation Postdoctoral Fellow.

\bibliographystyle{plain}
\bibliography{references}

\begin{thebibliography}{10}

\bibitem{ahnert2007numerical}
Karsten Ahnert and Markus Abel.
\newblock Numerical differentiation of experimental data: local versus global
  methods.
\newblock {\em Computer Physics Communications}, 177(10):764--774, 2007.

\bibitem{Benner2015siamreview}
Peter Benner, Serkan Gugercin, and Karen Willcox.
\newblock A survey of projection-based model reduction methods for parametric
  dynamical systems.
\newblock {\em SIAM review}, 57(4):483--531, 2015.

\bibitem{Billings2013book}
Stephen~A Billings.
\newblock {\em Nonlinear system identification: NARMAX methods in the time,
  frequency, and spatio-temporal domains}.
\newblock John Wiley \& Sons, 2013.

\bibitem{Bongard2007pnas}
J.~Bongard and H.~Lipson.
\newblock Automated reverse engineering of nonlinear dynamical systems.
\newblock {\em Proc. Natl. Acad. Sciences}, 104(24):9943--9948, 2007.

\bibitem{boninsegna2018sparse}
Lorenzo Boninsegna, Feliks N{\"u}ske, and Cecilia Clementi.
\newblock Sparse learning of stochastic dynamical equations.
\newblock {\em The Journal of chemical physics}, 148(24):241723, 2018.

\bibitem{Brunton2021koopman}
Steven~L Brunton, Marko Budi{\v{s}}i{\'c}, Eurika Kaiser, and J~Nathan Kutz.
\newblock Modern {K}oopman theory for dynamical systems.
\newblock {\em arXiv preprint arXiv:2102.12086}, 2021.

\bibitem{brunton2016pnas}
Steven~L. Brunton, Joshua~L. Proctor, and J.~Nathan Kutz.
\newblock Discovering governing equations from data by sparse identification of
  nonlinear dynamical systems.
\newblock {\em Proceedings of the National Academy of Sciences},
  113(15):3932--3937, 2016.

\bibitem{callaham2021role}
Jared~L Callaham, Steven~L Brunton, and Jean-Christophe Loiseau.
\newblock On the role of nonlinear correlations in reduced-order modeling.
\newblock {\em arXiv preprint arXiv:2106.02409}, 2021.

\bibitem{champion2020unified}
Kathleen Champion, Peng Zheng, Aleksandr~Y Aravkin, Steven~L Brunton, and
  J~Nathan Kutz.
\newblock A unified sparse optimization framework to learn parsimonious
  physics-informed models from data.
\newblock {\em IEEE Access}, 8:169259--169271, 2020.

\bibitem{chartrand2011numerical}
Rick Chartrand.
\newblock Numerical differentiation of noisy, nonsmooth data.
\newblock {\em International Scholarly Research Notices}, 2011, 2011.

\bibitem{de2020pysindy}
Brian {de Silva}, Kathleen Champion, Markus Quade, Jean-Christophe Loiseau,
  J~Nathan Kutz, and Steven Brunton.
\newblock Py{SIND}y: A {P}ython package for the sparse identification of
  nonlinear dynamical systems from data.
\newblock {\em Journal of Open Source Software}, 5(49):1--4, 2020.

\bibitem{delahunt2021toolkit}
Charles~B Delahunt and J~Nathan Kutz.
\newblock A toolkit for data-driven discovery of governing equations in
  high-noise regimes.
\newblock {\em arXiv preprint arXiv:2111.04870}, 2021.

\bibitem{fasel2021sindy}
Urban Fasel, Eurika Kaiser, J~Nathan Kutz, Bingni~W Brunton, and Steven~L
  Brunton.
\newblock {SIND}y with control: {A} tutorial.
\newblock {\em arXiv preprint arXiv:2108.13404}, 2021.

\bibitem{fasel2021ensemble}
Urban Fasel, J~Nathan Kutz, Bingni~W Brunton, and Steven~L Brunton.
\newblock Ensemble-{SIND}y: {R}obust sparse model discovery in the low-data,
  high-noise limit, with active learning and control.
\newblock {\em arXiv preprint arXiv:2111.10992}, 2021.

\bibitem{kaheman2020sindy}
Kadierdan Kaheman, J~Nathan Kutz, and Steven~L Brunton.
\newblock {SIND}y-{PI}: a robust algorithm for parallel implicit sparse
  identification of nonlinear dynamics.
\newblock {\em Proceedings of the Royal Society A}, 476(2242):20200279, 2020.

\bibitem{Kaiser2018prsa}
Eurika Kaiser, J~Nathan Kutz, and Steven~L Brunton.
\newblock Sparse identification of nonlinear dynamics for model predictive
  control in the low-data limit.
\newblock {\em Proceedings of the Royal Society of London A}, 474(2219), 2018.

\bibitem{kaptanoglu2021promoting}
Alan~A. Kaptanoglu, Jared~L. Callaham, Aleksandr Aravkin, Christopher~J.
  Hansen, and Steven~L. Brunton.
\newblock Promoting global stability in data-driven models of quadratic
  nonlinear dynamics.
\newblock {\em Phys. Rev. Fluids}, 6:094401, Sep 2021.

\bibitem{kaptanoglu2020physics}
Alan~A. Kaptanoglu, Kyle~D. Morgan, Chris~J. Hansen, and Steven~L. Brunton.
\newblock Physics-constrained, low-dimensional models for magnetohydrodynamics:
  First-principles and data-driven approaches.
\newblock {\em Phys. Rev. E}, 104:015206, Jul 2021.

\bibitem{Kutz2016book}
J.~N. Kutz, S.~L. Brunton, B.~W. Brunton, and J.~L. Proctor.
\newblock {\em Dynamic Mode Decomposition: Data-Driven Modeling of Complex
  Systems}.
\newblock SIAM, 2016.

\bibitem{Loiseau2017jfm}
J.-C. Loiseau and Steven~L. Brunton.
\newblock Constrained sparse {Galerkin} regression.
\newblock {\em Journal of Fluid Mechanics}, 838:42--67, 2018.

\bibitem{maddu2019stability}
Suryanarayana Maddu, Bevan~L Cheeseman, Ivo~F Sbalzarini, and Christian~L
  M{\"u}ller.
\newblock Stability selection enables robust learning of partial differential
  equations from limited noisy data.
\newblock {\em arXiv preprint arXiv:1907.07810}, 2019.

\bibitem{mangan2016inferring}
Niall~M Mangan, Steven~L Brunton, Joshua~L Proctor, and J~Nathan Kutz.
\newblock Inferring biological networks by sparse identification of nonlinear
  dynamics.
\newblock {\em IEEE Transactions on Molecular, Biological and Multi-Scale
  Communications}, 2(1):52--63, 2016.

\bibitem{messenger2021weakpde}
Daniel~A Messenger and David~M Bortz.
\newblock Weak {SIND}y for partial differential equations.
\newblock {\em Journal of Computational Physics}, page 110525, 2021.

\bibitem{pathak2018model}
Jaideep Pathak, Brian Hunt, Michelle Girvan, Zhixin Lu, and Edward Ott.
\newblock Model-free prediction of large spatiotemporally chaotic systems from
  data: a reservoir computing approach.
\newblock {\em Physical review letters}, 120(2):024102, 2018.

\bibitem{peherstorfer2016data}
Benjamin Peherstorfer and Karen Willcox.
\newblock Data-driven operator inference for nonintrusive projection-based
  model reduction.
\newblock {\em Computer Methods in Applied Mechanics and Engineering},
  306:196--215, 2016.

\bibitem{qian2020lift}
Elizabeth Qian, Boris Kramer, Benjamin Peherstorfer, and Karen Willcox.
\newblock Lift \& {L}earn: {P}hysics-informed machine learning for large-scale
  nonlinear dynamical systems.
\newblock {\em Physica D: Nonlinear Phenomena}, 406:132401, 2020.

\bibitem{markus_quade_sparsereg}
Markus Quade.
\newblock sparsereg - collection of modern sparse regression algorithms,
  February 2018.

\bibitem{Raissi2019jcp}
M~Raissi, P~Perdikaris, and GE~Karniadakis.
\newblock Physics-informed neural networks: A deep learning framework for
  solving forward and inverse problems involving nonlinear partial differential
  equations.
\newblock {\em Journal of Computational Physics}, 378:686--707, 2019.

\bibitem{raissi2017machine}
Maziar Raissi, Paris Perdikaris, and George~Em Karniadakis.
\newblock Machine learning of linear differential equations using {G}aussian
  processes.
\newblock {\em Journal of Computational Physics}, 348:683--693, 2017.

\bibitem{Reinbold2020pre}
Patrick~AK Reinbold, Daniel~R Gurevich, and Roman~O Grigoriev.
\newblock Using noisy or incomplete data to discover models of spatiotemporal
  dynamics.
\newblock {\em Physical Review E}, 101(1):010203, 2020.

\bibitem{reinbold2021robust}
Patrick~AK Reinbold, Logan~M Kageorge, Michael~F Schatz, and Roman~O Grigoriev.
\newblock Robust learning from noisy, incomplete, high-dimensional experimental
  data via physically constrained symbolic regression.
\newblock {\em Nature communications}, 12(1):1--8, 2021.

\bibitem{Rudy2017sciadv}
Samuel~H Rudy, Steven~L. Brunton, Joshua~L. Proctor, and J.~Nathan Kutz.
\newblock Data-driven discovery of partial differential equations.
\newblock {\em Science Advances}, 3(e1602614), 2017.

\bibitem{Schaeffer2017prsa}
Hayden Schaeffer.
\newblock Learning partial differential equations via data discovery and sparse
  optimization.
\newblock In {\em Proceedings of the Royal Society A}, volume 473, page
  20160446. The Royal Society, 2017.

\bibitem{Schaeffer2017pre}
Hayden Schaeffer and Scott~G McCalla.
\newblock Sparse model selection via integral terms.
\newblock {\em Physical Review E}, 96(2):023302, 2017.

\bibitem{schmid2010dynamic}
Peter~J Schmid.
\newblock Dynamic mode decomposition of numerical and experimental data.
\newblock {\em Journal of fluid mechanics}, 656:5--28, 2010.

\bibitem{schmidt_distilling_2009}
Michael Schmidt and Hod Lipson.
\newblock Distilling free-form natural laws from experimental data.
\newblock {\em Science}, 324(5923):81--85, 2009.

\bibitem{tibshirani2011solution}
Ryan~J Tibshirani and Jonathan Taylor.
\newblock The solution path of the generalized lasso.
\newblock {\em The annals of statistics}, 39(3):1335--1371, 2011.

\bibitem{vlachas2018data}
Pantelis~R Vlachas, Wonmin Byeon, Zhong~Y Wan, Themistoklis~P Sapsis, and
  Petros Koumoutsakos.
\newblock Data-driven forecasting of high-dimensional chaotic systems with long
  short-term memory networks.
\newblock {\em Proc. R. Soc. A}, 474(2213):20170844, 2018.

\end{thebibliography}

\end{document}